# Self Phase Modulation and Cross Phase Modulation in Nonlinear Silicon Waveguides for On-Chip Optical Networks - A Tutorial

Abdurrahman Javid Shaikh[1, 2, a] *, Othman Sidek[1, b] and Fauzi Packeer[1, c]

[1] School of Electrical and Electronic Engineering, Engineering Campus, Universiti Sains Malaysia, 14300 Nibong Tebal, Seberang Perai Selatan, Pulau Pinang, Malaysia.

[2] Department of Electrical Engineering, NED University of Engineering & Technology, Karachi - 74270, Pakistan.

[a]arjs@neduet.edu.pk, [b]othman.cedec@usm.my, [c]fauzi.packeer@usm.my

* Corresponding author

**Abstract.** Silicon is a nonlinear material and optics based on silicon makes use of these nonlinearities to realize various functionalities required for on-chip communications. This article describes foundations of these nonlinearities in silicon at length. Particularly, self phase modulation and cross phase modulation in the context of integrated on-board and on-chip communications are presented. Important published results and principles of working of these nonlinearities are presented in considerable detail for non-expert readers.

**Introduction**

The ultimate goal of any communication system design, whether long-haul or short reach, is to maximize data transmission capacity while maintaining a reasonable level of design and operation cost. The long-haul communication system involves signal transmission and hence data communication through atmosphere via wireless medium or under the sea through fiber optic cables. The short reach data communication, on the other hand, involves signal transmission between two points much less distant. These short reach interconnects are common in big data centers and supercomputing facilities. However, since the inception of multicore processors and multiprocessors systems, the focus of research in the field of interconnect has been shifted primarily towards interconnects of even smaller scale, i.e. intrachip and interchip interconnects. This article will specifically discuss physical processes and their applications in the context of these small scale interconnects.

It is well known that speed of a microprocessor is much faster as compared to the transmission speeds of chipscale interconnects. These slow metallic interconnects, which are responsible for data exchange between processor and memory, result in a situation where a single-core CPU remains mostly idle during the read/write cycles. One can imagine the undesirability of the situation that would arise when multicores are involved. Contemporary multiprocessors system can simply outpace the interconnect transmission speeds and hence are severely underutilized. This underutilization has become more relevant since the paradigm of *big data* got widespread acceptance. The idea of big data relies solely on extracting conclusions and making inferences by simultaneously processing massive data stored in data centers rather than dividing it into smaller subgroups. This data (may be in exabytes or even zettabytes in future) are now not limited to any exotic scientific research application anymore. The applications may include anything from behavioral analysis to predicting market trends to forecasting natural disasters with unprecedented precisions. Hence the applications are ubiquitous in nature and realization requires massively parallel processing capabilities.

## The Envisioned Solution

Short reach optical communications has been one of the most sought after research fields to improve interconnect and hence device and system performance [1]. The interest in optical interconnects arise from a number of superior characteristics displayed by these interconnects as opposed to their electronic counterparts. Firstly, optical interconnects tightly confine the electromagnetic fields at much higher frequency of operation hence result in less interference and higher device integration densities. The authors' work on the analysis and design of such high confining, single mode rib waveguides at nanoscale provides detailed design guidelines for the CMOS-compatible silicon-on-insulator (SOI) material system [2]. The SOI platform has proven to be successful in the photonics research arena for more than a decade now, and is further explored for novel applications involving spintronics [3], negative differential capacitance [4], and silicon rich oxides [5] to name a few amongst others [6]. Secondly, optical interconnects are much less power hungry as compared to metallic interconnects. According to the 2007 *International Technology Roadmap for Semiconductors (ITRS)* report, 51% of the total processor power (at 130nm node) is consumed by metallic interconnects [7]. This percentage is bound to increase substantially as the gate length would shrink if the design philosophy remains the same. Consequently, optical interconnects which dissipate much smaller amounts of power are also promising to keep a reasonable running cost for data centers. According to *Greenpeace* report 2010, in 2007 alone, the energy consumption of all data centres amounted to 623 billion KWh [8]. Precise consumption by interconnects is yet to be determined. Nevertheless, considering aggregate length of interconnects involved in a data centre and the resultant amount of heat dissipation coupled with power consumption in cooling systems, it can be safely concluded that most of the energy in these centres is consumed in heat dissipation and heat management systems.

To enhance transmission capacity and avoid effects of electrical parasitics, the system envisioned is based on chipscale wavelength division multiplexing (WDM) network which utilizes multiple closely spaced wavelengths, each acting as a channel for simultaneous signal transmission. The bandwidth of each channel is essentially limited by data modulation speed for that channel. Modulation speed, on the other hand, is limited by electrical parasitics (resistance and capacitance) associated with modulator's metallic electrodes, and mobilities of electrons and holes in the material [9]. Nevertheless, silicon based optical modulator as fast as 50Gbps (giga-bits-per-second) has been demonstrated [10-12]. Now, for example, a 16-channel WDM transceiver where each channel operating at 50Gbps can process signals with an aggregate speed of 800Gbps (16 channels x 50Gbps/channel) over a single optical connection. Achieving this speed in contemporary metallic short reach interconnects is far from possible.

In the light of above discourse one might rightly conclude that if modulation bandwidth per channel is limited then total transmission capacity can still be enhanced by increasing number of channels. However, maximum number of usable channels is limited by optical impairments of a communication medium. At low to moderate optical intensities (determined by signal amplitude) intrinsic maximum speed of an optical interconnect is fundamentally limited by linear optical impairments which are characterized by pulse dispersion. Digital signals, in the form of light pulses, sent down an optical interconnection are broadened and lose their amplitude. This broadening effect would occur even in an otherwise attenuation free medium. If left unmanaged, pulse dispersion will cause inter-symbol-interference beyond its resolvable length (Fig. 1).

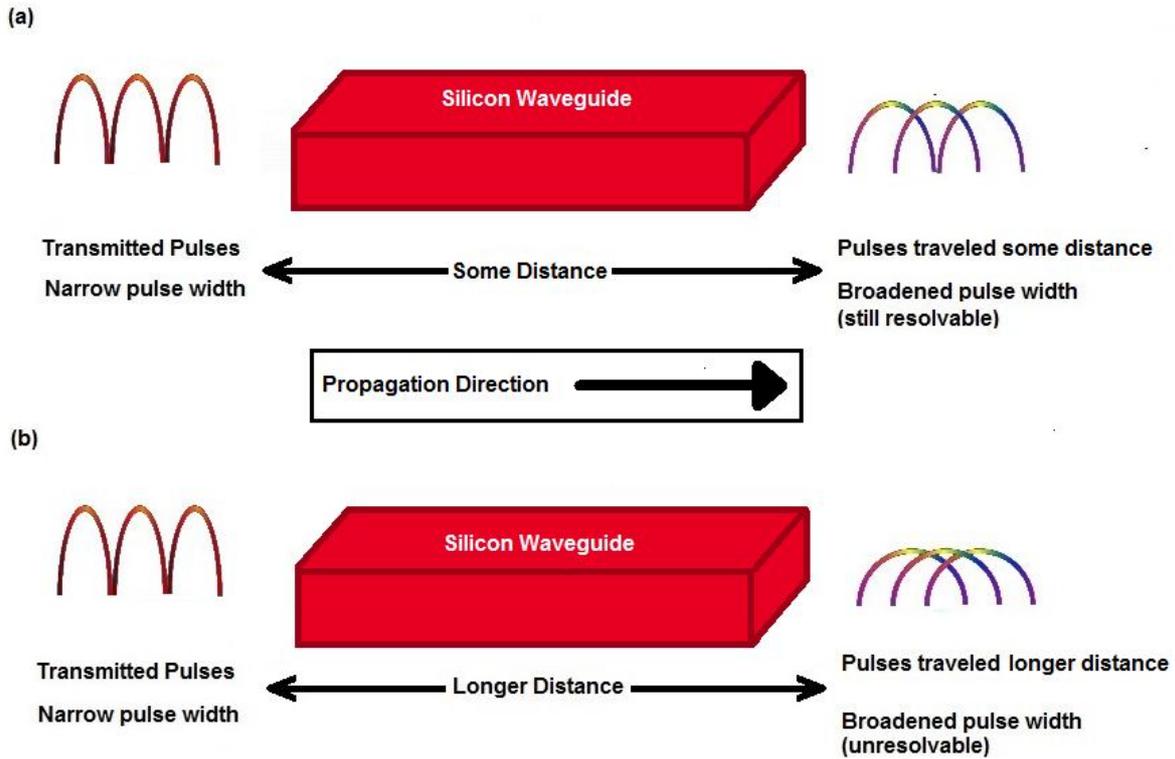

Figure 1. Pulse dispersion, where narrow input pulses are broadened by linear optical impairments, rendering the output pulses of smaller amplitude, (a) resolvable after travelling a certain distance, or (b) unresolvable after travelling beyond resolvable length.

To realize an efficient WDM system on chip, a multitude of optical functionalities are required to be demonstrated. Many of the components in silicon like third harmonic source [13-14], stimulated Raman scattering (SRS) based mid-IR amplifiers and lasers [15-23]; and processes like wavelength conversion [24-27], sensing [28-29], and pulse shaping [30-31] have already been demonstrated. Most of these components are based on nonlinearities associated with the silicon material. *Self Phase Modulation (SPM)* and *Cross Phase Modulation (XPM)* are two of such nonlinearities in silicon studied in quite detail in literature and used in applications pertaining to optical communications and computing. Whether it is SPM, XPM or any other nonlinear behaviour, all are attributed to nonlinear polarization of a material with respect to applied field. In the following section we explained basis of nonlinearity in silicon before taking SPM and XPM for discussion.

**Linear and nonlinear silicon photonics**

Silicon became material of choice because of its favourable material properties like high refractive index (~3.47 at 1550 nm), high thermal conductivity (1.56W/cm/K), and extremely low intrinsic absorption loss at technologically important wavelengths (1310 nm and 1550 nm) coupled with extensive material knowledge and highly mature fabrication infrastructure. It is the ability to pattern and fabricate nanoscale features and devices in silicon which gave a fresh impetus to the field of nonlinear silicon photonics. Critical dimension control of contemporary fabrication facilities enables precise control of waveguide dimensions. Such control on fabrication parameters enabled realization of photonic bandgap materials and other simpler yet compact structures based on rib and slot waveguides. This ability also allowed fabrication of the so called left-handed (or negative index) materials out of both silicon-metal structure [32] and other materials [33-36], enabling features which were not possible otherwise. Waveguide devices now have feature sizes of few tens of nanometers as compared to several hundred nanometers in 1990s. Modern fabrication technology has achieved substantially smaller device sizes for micro and nanoelectronics applications, however, with current knowledge in optical devices, such small guiding structures means unacceptably high insertion losses if used in photonic applications. High refractive index of silicon coupled with cladding of low-index (~1.45) enables strong confinement of light which, in turn, enables photonic device design using ultrasmall cross sections waveguides for various applications.

Nonlinear optical processes are observed when dielectric or semiconducting material is irradiated by sufficiently high intensity electromagnetic field. This field interacts with electrons and crystal lattice of the material which engenders nonlinear processes. The changing of energy states of charges inside a material with respect to electric component of applied optical field "activates" the electronic component of nonlinearity, known as Kerr effect or Kerr nonlinearity. On the other hand, interaction of the same field with vibration state of crystal lattice gives rise to phonon-assisted nonlinear processes like Brillioun scattering (involving acoustic-phonon) and Raman scattering (involving optical-phonon). Phonons are quasiparticles which relate to vibrational excitation of a solid and define many of important characteristics of a material like thermal and electrical conductivities. The change in state (electronic and vibrational) manifests itself as polarization (**P**) of the material due to the applied field (**E**), that can be given by power series of **E**:

$$\mathbf{P}(t) = \varepsilon_0 (\chi^{(1)} \mathbf{E}(t) + \chi^{(2)} \mathbf{E}^2(t) + \chi^{(3)} \mathbf{E}^3(t) + \cdots), \tag{1}$$

where, $\varepsilon_0$ is the permittivity of free space and $\chi^{(1)}$ is the first order optical susceptibility, while $\chi^{(2)}$ and $\chi^{(3)}$ are the nonlinear second and third order susceptibilities and so on. In silicon, nonlinear processes originated by susceptibilities of first three orders have been widely studied [37-38]. For isotropic materials – properties of which are independent of the direction of excitation field (signal) – susceptibilities are scalar quantities. However, for anisotropic materials susceptibilities are tensors of various ranks. Silicon's first-order susceptibility is isotropic, having a single complex value, while third-order term is a tensor of rank 4 due to its crystalline nature [38]. The second order susceptibility term ($\chi^{(2)}$) is absent in silicon understandably for its highly symmetrical (centrosymmetric) structure. This absence is far from being desirable as this term is required for realization of high speed electro-optic modulators. Methods have been applied to circumvent this shortcoming of silicon by applying molecular stress; however, discussing this method is beyond the scope of this article.

**Linear Complex Refractive Index and Pulse Dispersion.** The first order susceptibility term ($\chi^{(1)}$) controls the linear complex refractive index of a material and contains two major contributions ($\chi_L^{(1)}$ and $\chi_D^{(1)}$). One, from bound charges of material described by Lorentz model ($\chi_L^{(1)}$), and the other from free carriers (electrons and holes) in the material, described by Drude model ($\chi_D^{(1)}$). Since the bound charges (which are bound to their parent nuclei in the form of charge cloud) cannot drift freely under the influence of applied (electric component of) optical field, they tend to oscillate with the time-varying field. The Lorentz model is therefore a frequency (or wavelength) dependent model based on classical spring-mass oscillator system widely used in mechanics. The Lorentz contribution to susceptibility is therefore expected to manifest dramatic variation in susceptibility near some resonance point where absorption loss is maximum. Fortunately, in the case of silicon, this point is well away from the technologically important wavelengths. Drude's model, on the other hand, is free from any resonance effects as free carriers in a material can drift swiftly under the influence of applied field and are negligibly bound to positively charged nucleus. Although, its value changes based upon wavelength of operation, $\chi_D^{(1)}$ reflects only the effect of free carriers on linear complex refractive index ($n_2$). Hence, the complex linear refractive index is given by:

$$\boldsymbol{n_1} = 1 + \boldsymbol{\chi}_L^{(1)} + \boldsymbol{\chi}_D^{(1)}. \tag{2}$$

For silicon, more accurate empirical formulae based on extensive experimental data have been developed for telecommunication wavelengths [39-41]. These formulae are widely used to estimate phase shifts and absorption losses in optical modulators based on carrier density variations in waveguides.

The dispersion contributions to total pulse dispersion originate from the wavelength dependence of linear refractive index term (specifically from $\boldsymbol{\chi}_L^{(1)}$). Pulse dispersion includes the effects of *material dispersion*, *intermodal dispersion*, *waveguide dispersion* and *polarization mode dispersion*. The *material dispersion* originates from the fact that refractive index of any transmission medium depends on the wavelength of operation. Different component wavelengths of an input pulse would travel at different velocities, causing the pulse to broaden in time-domain as it propagates. Optical energy propagates down any medium in the form of electromagnetic modes supported by the geometric structure of the medium. A wave-guiding structure may support only fundamental mode or may support multiple modes (may be dozens) depending on the size of the core relative to the wavelength of operation, cladding material and exact cross-section of the waveguide. *Intermodal dispersion* occurs because different modes supported by the waveguide (WG), at a particular wavelength, travel with varying longitudinal velocities because of different (yet defined) path taken by each supported mode while propagating. This mechanism, again, leads to temporal pulse broadening. However, chipscale interconnects are expected to support fundamental mode only (like single mode fibers). Therefore, for single mode optical waveguides, the dominant contribution in pulse dispersion comes from the effect of waveguide geometry on transmission characteristics of the fundamental mode. It should be emphasized that material dispersion is important primarily in bulk materials where plane wave approximation holds. In waveguides, however, transmission properties still depend on wavelength of operation even if material dispersion is neglected altogether. This is called waveguide dispersion which is used to engineer total dispersion of a guiding medium. In practice, material dispersion is incorporated in calculating waveguide dispersion using Sellmeier equation (see, for example, Eq. (1) of [38]). increase the channel capacity of a WDM system, one must manage pulse dispersion to keep it to the minimum possible value. In long-haul optical communication system pulse dispersion is effectively managed by dispersion compensating fibers. Using the fact that different contributions to pulse dispersion bear different signs (positive or negative), total dispersion can be minimized. Same dispersion engineering has been applied in photonic waveguides to demonstrate management of total dispersion (which is inversely related to maximum achievable bit rate in a medium) [42-45].

Having discussed wavelength and free carrier dependence of the refractive index and its relation with first-order susceptibility, $\chi^{(1)}$, and hence the linear polarization; we can consider the wavelength dependence of propagation constant (β) which quantifies phase change of propagating pulse with respect to the distance traveled. Various order derivatives of the propagation constant with respect to frequency (or wavelength) constitute corresponding orders of linear dispersion. The first order derivative (denoted by $β_1$) represents group velocity in the propagation of overall signal envelope and does not affect its shape. The second order derivative ($β_2$) represents group velocity dispersion (GVD) which affects temporal shape of the envelope and causes the pulse to broaden. This broadening reduces the transmission capacity of the optical interconnects at particular wavelength. Due to the tight optical confinement and subwavelength dimensions of silicon waveguides, the dispersion characteristics are far more profound as compared to silica fibers and large cross section waveguides, rendering considerable pulse broadening in chipscale interconnects [38]. Moreover, this tight confinement and small size allow the waveguide cross section to have decisive effect on overall dispersion. Using the precision in fabrication offered by silicon material system, GVD can be nullified to achieve zero-GVD (ZGVD) point at the wavelength of interest by adjusting waveguide width and height (and etch depth in case of rib waveguides) [42, 45]. Higher-order dispersion, like third order dispersion (TOD), fourth order dispersion, and so on are similarly defined. TOD has significant affect on pulse profile in the case of SPM while other high order dispersion parameters do not affect pulse broadening with considerable degree.

**Non-Linear Complex Refractive Index.** At sufficiently high optical fields, the third order susceptibility ($\chi^{(3)}$) becomes significant and the behavior of silicon is now influenced by both the linear and nonlinear susceptibilities. Like $\chi^{(1)}$, whose real part is related to real part of linear refractive index (n) and imaginary part is linear absorption loss (α); the real part of $\chi^{(3)}$ represents nonlinear refractive index ($n_2 I$; $n_2 \rightarrow$ Kerr coefficient, I $\rightarrow$ optical intensity) while the imaginary part is nonlinear two-photon absorption coefficient ($\alpha_2 \rightarrow$ TPA coefficient). Kerr effect (quantified by the coefficient $n_2$) is an ultrafast response of electrons to the applied field and is therefore considered instantaneous; so is two-photon absorption. TPA is a frequency dependent nonlinear effect where two photons from high intensity pulses are absorbed in a nonlinear medium such that a bound charge breaks away from the parent nucleus and becomes free. This effect represents a loss mechanism limiting the maximum speed that can be observed in a device based on any nonlinear process. For example, free carriers in silicon have lifetimes in the range of few to several nanoseconds as opposed to the ultrafast processes engendered by the Kerr effect whose response times are in femtoseconds. These free carriers increase the relaxation time and hence limit the maximum bit rate on which the nonlinear device may work. Moreover, free carriers distort real part of linear refractive index (known as *free carrier index* (FCI)) and increases linear absorption (known as *free carrier absorption* (FCA)), as dictated by the Drude model discussed above. TPA augments generation of free carriers which limits device overall speed and increases FCA loss.

For every medium which can exhibit Kerr nonlinearity (Kerr medium), usually a wavelength dependent nonlinear figure of merit (FOM) is defined to quantify relative values of Kerr and TPA coefficients in that medium. Higher Kerr and lower TPA values would give higher values of the FOM. While the Kerr coefficient is significant in silicon, the TPA coefficient is such that the overall FOM is well below unity. For nonlinear devices to be efficient, one of the requirements is to have FOM>2. The nonlinear complex refractive index ($\mathbf{n_2}$) is intensity dependent, and is significant at high field intensities. This does not mean that $\mathbf{n_2}$ does not show frequency dependence. In fact, dispersion of both (Kerr and TPA coefficients) in silicon has been studied extensively [46-49].

$\chi^{(3)}$ term represents nonlinear phenomena like stimulated Raman scattering, self phase modulation, four-wave mixing (FWM), cross phase modulation and third harmonic generation (THG). Not discussed in this article, however, that $\chi^{(3)}$ has another contribution, besides Kerr and TPA, that is the Raman response. For silicon, the Raman contribution is negligible in the near infrared regime and $\chi^{(3)}$ represents only SPM, TPA, XPM, THG, and FWM processes.

The nonlinear behavior of silicon can be enhanced (i) by making the optical field travel slow inside waveguides (slow light), and (ii) by increasing optical field confinement. The former increases probability of photon interaction with WG crystal while later enhances the electric field inside the WG. Increased electric field inside WG means increased energy density which, in turn, means occurrence of nonlinear phenomena at much reduced input power (i.e., low-threshold). $\chi^{(3)}$ in silicon is already 3-4 orders of magnitude higher than silica, moreover, tight optical confinement further enhances effective value of this susceptibility. It must be noted, however, that higher values of effective susceptibility is undesirable in optical interconnects scenario. High $\chi^{(3)}$ will broaden pulse through SPM or may cause interchannel crosstalk, prohibiting closely spaced channels in WDM based systems and hence reducing transmission capacity. Still, as detailed in next two sections, high nonlinearity of silicon can be utilized to our advantage enabling crucial functionalities required for realization of chipscale WDM networks.

## Self Phase Modulation (SPM)

SPM is an intensity dependent phenomenon wherein change in nonlinear refractive index causes a pulse propagating through a waveguide to broaden its spectral width while the temporal profile largely remains unchanged. An ultrashort high intensity pulse induces time and position-varying change in local refractive index of the material causing phase shift in the propagating pulse. In other words, the constituent spectral components of the pulse bearing specific amount of energy in each component would induce an specific local index change in the medium while propagating through it. This change in index would affect the spectrum of the pulse by "generating" more frequency components at the expense of power in the original spectral components.

SPM has been used to demonstrate supercontinuum generation where a femtoseconds pulse is broadened in spectral domain at the expense of pulse power. For example, Fig. 2 presents: (a) temporal, while (b) presents spectral normalized intensity profiles (unitless) of a femtoseconds pulse propagating in a ZGVD strip-guide centered at 1550 nm [43]. Increasing the pulse power increases this wavelength domain broadening while SPM does not distort the symmetry of the envelope in both domains. However, TOD distorts this temporal and spectral symmetry and adds time domain oscillations in the profile of the pulse as shown by Fig. 2. Like ZGVD, TOD in silicon has also been demonstrated to show zero dispersion by adjusting waveguide dimensions [38]. SPM together with free carrier induced refractive index change can augment spectral broadening as high as >30x. However, it is TPA which defines the upper limit in the pulse broadening as the TPA increases with pump source intensities as does the overall pulse broadening. Spectral broadening in silicon waveguides cascaded with microdisk/microring wavelength filters [50] can be used for channel wavelength generation as shown in Fig. 3.

Hsieh and coworker [51] used a silicon waveguide with rectangular cross section that was pumped with a laser system (1300 nm-1600 nm) with half power bandwidth of ~30 nm and a repetition rate of 250 kHz. When ultrashort pulses of width 100 fs at a wavelength of 1310 nm were pumped, a pulse broadening from ~80nm to more than 350nm was observed depending upon coupled pump intensities. Fig. 4(a) shows reported spectral width broadening for various pump powers while 4(b) shows experimental and simulated dependence of spectral width on couple peak power [51].

Koonath et al. [50], went a step further and fabricated microdisk resonators to slice and drop several wavelengths on the same chip. This demonstration may result in an aforementioned much needed, low cost, chipscale WDM source for inter/intrachip optical communications. For silicon based WDM on chip networks to work reliably, high pumps may not be suitable for it may degrade material properties over time and would dissipate excess heat. Lately, continuum generation of 500nm spectral width with 5 picoJoules pulses have been demonstrated in hydrogenated amorphous silicon (a-Si:H) waveguides [52]. a-Si:H also performed quite stable as a material when exposed to high power density pulses for several hours. In silicon, peak of Kerr coefficient occurs beyond source wavelength of 1850nm where TPA is negligible resulting in high FOM values [46]. Using this regime, generation of continuous band of 1000 nm width has been reported with picoseconds pulses centered at 1950 nm [53]. Such demonstrations are very promising for reliable WDM sources where linear dispersion management coupled with supercontinuum generation and wavelength filters are potential and cost effective replacements of heat dissipating multiple lasers topology.

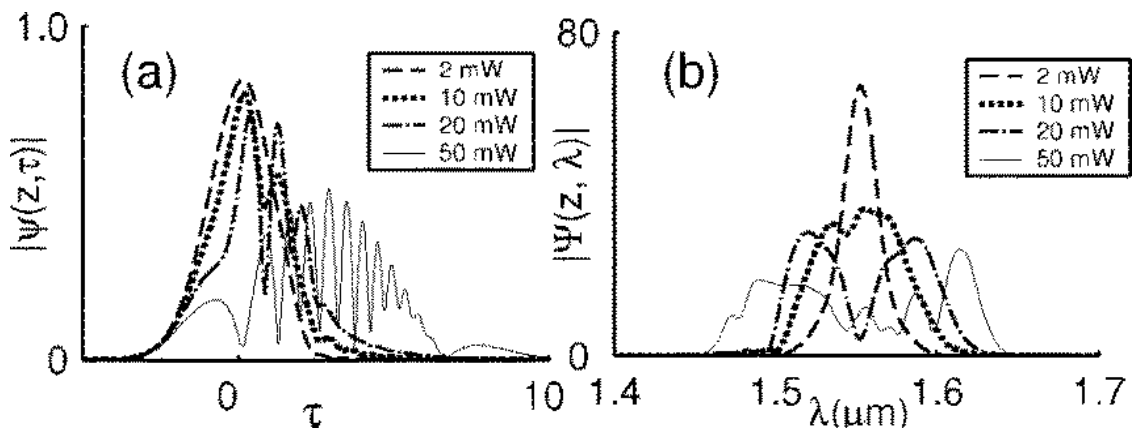

Figure 2. (a) Temporal and (b) Spectral normalized intensity profiles of a femtosecond pulse centered at 1550nm, propagating through dispersion engineered silicon waveguide with ZGVD wavelength at 1550nm. The profiles are shown for different pulse powers. $\tau$ is the normalized time $(T/T_0)$ where $T_0$ is pulse width. Copyright © 2006 IEEE. All rights reserved. Reprinted, with permission, from [43].

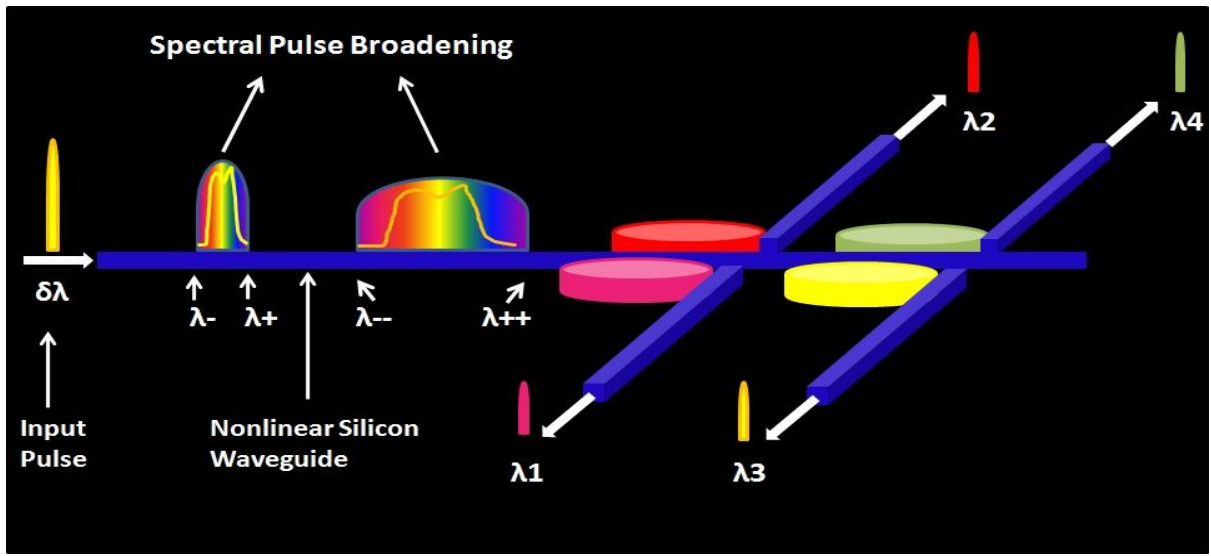

Figure 3. Concept diagram of WDM optical source. A narrow linewidth input pulse gets progressively broadened while traveling through nonlinear silicon waveguide due to SPM. Four microdisk resonator filters of different radii pick a pulse signal of particular wavelength.

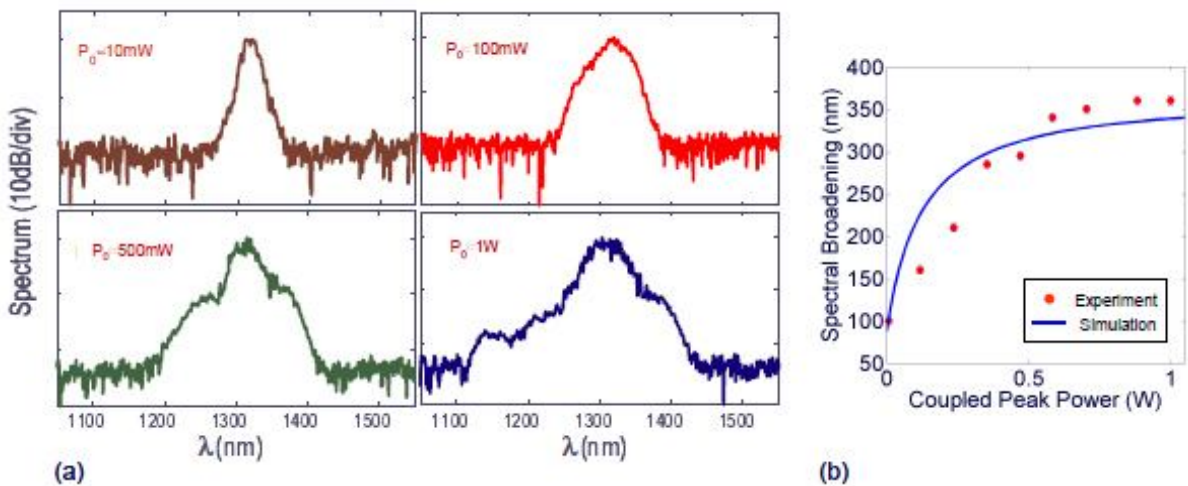

Figure 4. (a) Spectral width broadening vs various pump powers. (b) Experimental and simulated dependence of spectral width on couple peak power. Copyright © 2007 OSA. All rights reserved. Reprinted, with permission, from [51].

**Cross phase modulation (XPM)**

In a WDM link where several closely spaced channels traverse the same medium, XPM can cause channel crosstalk. When an optical field, comprising at least two nearly monochromatic waves copropagates in a Kerr medium, the field intensity of one can be shown to induce nonlinear refractive index change for the other and vice versa, known as cross-phase modulation. Consider, an optical field (E) comprising two monochromatic waves ($E_1$ and $E_2$) of frequencies $\omega_1$ and $\omega_2$ such that $E = E_1 + E_2 = Re[E(\omega_1)e^{j\omega_1 t} + E(\omega_2)e^{j\omega_2 t}]$; it can be shown that Eq. (1) contains the following term in its simplest form, representing the cross phase modulation effect;

$$\mathbf{P}(t)^{XPM} = \varepsilon_0 \chi^{(3)} |E(\omega_1)||E(\omega_2)| [E(\omega_1)cos\omega_2 t + E(\omega_2)cos\omega_1 t] \quad (3)$$

In a WDM system, hence, phase of one channel will eventually be associated with intensity of the other. It can also be shown mathematically that the induced nonlinear index is twice the magnitude of SPM induced index. For multiple data modulated channels where the monochromatic nature is lost, the XPM interactions are more complicated and each wavelength is influenced by intensity of every other spectral component of the pulse.

XPM in silicon has been studied in depth at 1550 nm and has been utilized to demonstrate functionalities like optical switching, format conversion and wavelength conversion. An initial report of experimental XPM based all-optical interferometric switch in silicon-on-insulator (SOI) WGs using Mach-Zehnder interferometer (MZI) configuration was made in 2004 [54]. The results demonstrated switching of continuous wave (CW) signal ~25nm away from pump laser. Optical switches are instrumental in WDM network architecture irrespective of the scale and are used for switching and routing multiplexed data. Fig. 5 shows the block diagram of the configuration. A modelocked fiber laser (MFL) at wavelength ($\lambda_p$) of 1560nm and pulse repetition rate of 20MHz, with pulses of width <1ps, was used as pump and a cw diode laser at wavelength ($\lambda_s$) of 1537nm was used to generate probe signal. The polarization beam splitter (PBS) was adjusted so that the pump pulses were routed towards the lower path where SOI WG is present and probe signal was split between the upper and lower paths of MZI. A retro-reflector in the upper path of MZI was used to tune relative signal delay between the two paths. The reflector was adjusted so that there was no signal at the output when there was no pulse in the lower path. Now as the sub-picosecond, high intensity pump pulse copropagates in the lower arm, it induces a nonlinear index in the WG and hence a phase shift is experienced by the CW probe in the lower arm. The two signals were combined in 50/50 splitter so that interference occurs among each spectral component of the signals. The switched signal was then recorded by a realtime digitizing oscilloscope after filtering pump out of the MZI output. This ultrafast switching occurs by virtue of XPM-induced phase difference between the two arms.

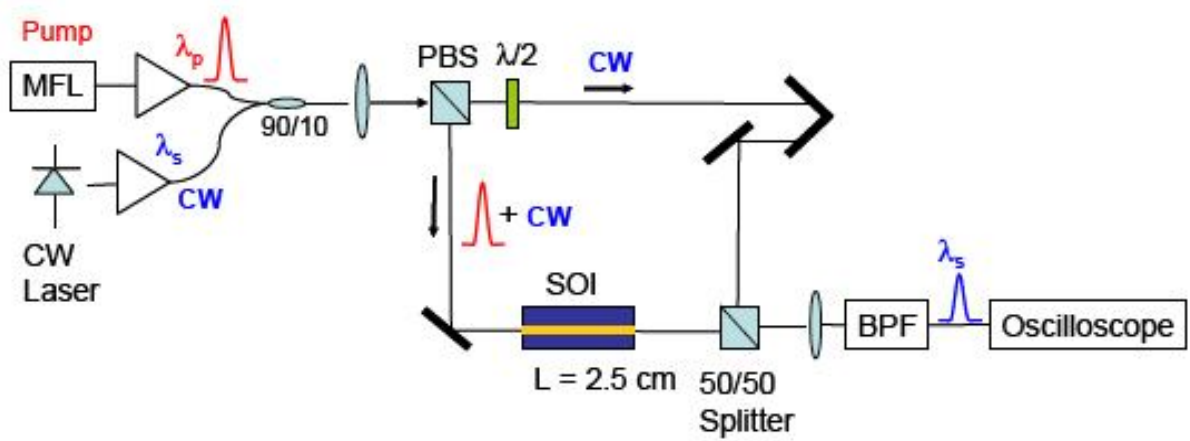

Figure 5. "Experimental setup of XPM based silicon switch. Mach Zehnder interferometer is used for switching. XPM induced phase shift causes switching of CW signal to the output port." Copyright © 2004 OSA. All rights reserved. Reprinted, with permission, from [54].

Ultrafast switching requires that recombination lifetime of the free carriers must be similar or shorter than the pump pulse width. However this is difficult to achieve for the pump produces picosecond pulses, while in silicon WGs the effective lifetime (~1ns) is much longer than the optical pulse width. This produces TPA induced FCA inside the active region of the WG, increasing device loss and relaxation time. A reversed biased *pin* waveguide structure may be used as a possible solution to actively remove free carriers generated due to TPA. However, use of pumped intensity equal to or lower than ~1 GW/cm$^2$ seemed to have solved the problem [54]. In this region free carrier generation is negligible while Kerr induced phase shift of 180$^o$ can be achieved. Other XPM based high speed switching has also been demonstrated more recently based on microring structures [55] as opposed to MZI configuration [54].

Another application of XPM which is very promising as a chipscale device is format conversion. Non-return-zero (NRZ) to RZ format conversion is attractive for many advantages of RZ over NRZ format, like, increased receiver sensitivity and transmission distances. XPM based format conversion requires phase modulation of a data (probe) signal (NRZ) by a pump signal (RZ), and then the phase modulated signal is transformed into amplitude modulated signal through a filter which is slightly detuned as compared to the probe's carrier. NRZ-OOK (On-Off Keying) to RZ-OOK format conversion at 10Gbps has been demonstrated in a 5mm long silicon nanowire by Astar et al. for the first time in 2009 [56]. Unlike using other nonlinear technique (like FWM), XPM format conversion is near the wavelength of that of the original NRZ signal.

A typical WDM network is classified as wavelength-routed network which routes data according to the channel's wavelength. To handle any congestion at any node within an onchip optical network, it is imperative to allocate available channel wavelength to incoming stream. Hence versatile wavelength conversion in onchip WDM networks is as essential as they are in optical fiber networks. XPM wavelength conversion is preferable for its insensitivity to dispersion, unlike FWM wavelength conversion where dispersion is detrimental, limiting spectral range where conversion is possible. Tuneable wavelength conversion of 10 Gbps RZ-OOK signal in silicon nanowire has been shown using XPM [57]. Demonstrated wavelength conversion was tuneable over a range of 20nm. More recently, error-free wavelength conversion of RZ signal with a tremendous speed of 160Gbps has also been demonstrated [58]. This demonstration together with experimental demonstrations of silicon based chipscale wavelength selective WDM filters are truly enabling for realization of high speed, congestion managed, onchip WDM networks.

**Conclusions**

In this paper, we discussed origin of linear and nonlinear behavior of silicon at length and showed how nonlinearities like self phase modulation and cross phase modulation can be utilized for realization of high speed WDM networks onchip. We discussed significant results reported using these nonlinearities to realize optical functionalities. Encouraging nonlinear optical characteristics of silicon, high field confinement and high optical damage threshold made these applications possible while availability of high quality silicon-on-insulator waveguides, high thermal conductivity, and ultrasmall feature size made these devices compatible for mass manufacturing. Silicon based optical communications has the potential to revolutionize computing by providing huge bandwidth and signal propagation with unprecedented speed where silicon's nonlinear behavior enables realization of diverse functionalities which were not possible before.

## Acknowledgement

The authors would like to acknowledge the support provided by the USM Fellowship scheme of Universiti Sains Malaysia.
## References

[1] D. A. B. Miller, Device Requirements for Optical Interconnects to Silicon Chips, Proceedings of the IEEE 97 (2009) 1166-1185.

[2] A. J. Shaikh and O. Sidek, Holistic Analysis and Systematic Design of High Confinement Factor, Single Mode, Nanophotonic Silicon-on-Insulator Rib Waveguides, Journal of Nanoelectronics and Optoelectronics (accepted for publication) (2016)

[3] J. Ghosh, *et al.*, Enhancement of Electron Spin Relaxation Time in Thin SOI Films by Spin Injection Orientation and Uniaxial Stress Journal of Nano Research 39 (2016) 34-42.

[4] A. Evtukh, *et al.*, Capacitive Properties of MIS Structures with SiOx and SixOyNz Films Containing Si Nanoclusters, Journal of Nano Research 39 (2016) 162-168.

[5] A. Kizjak, *et al.*, Electron Transport through Thin SiO2 Films Containing Si Nanoclusters, Journal of Nano Research 39 (2016) 169-177

[6] I. T. Kogut, *et al.*, The Device-Technological Simulation of Local 3D SOI-Structures Journal of Nano Research 39 (2016) 228-234

[7] ITRS. (2007, The International Technology Roadmap For Semiconductors 2007. *(June, 2011)*, 42. Available: http://www.itrs.net/links/2007itrs/execsum2007.pdf

[8] G. International, "Make IT Green: Cloud computing and its contribution to climate change," Greenpeace International, AmsterdamMarch 30, 2010 2010.

[9] G. T. Reed, *et al.*, Recent breakthroughs in carrier depletion based silicon optical modulators, Nanophotonics 3 (2013) 229–245.

[10] D. J. Thomson, *et al.*, 50-Gb/s Silicon Optical Modulator, IEEE Photonics Technology Letters 24 (2012) 234-236.

[11] T. Baba, *et al.*, 50-Gb/s ring-resonator-based silicon modulator, Optics Express 21 (2013) 11869-11876.

[12] X. Tu, *et al.*, 50-Gb/s silicon optical modulator with traveling-wave electrodes, Optics Express 21 (2013) 12776-12782.

[13] A. J. Shaikh and O. Sidek, Making Silicon Emit Light Using Third Harmonic Generation, Procedia Engineering 29 (2012) 1456-1461.

[14] CorcoranB, *et al.*, Green light emission in silicon through slow-light enhanced third-harmonic generation in photonic-crystal waveguides, Nat Photon 3 (2009) 206-210.

[15] R. Jones, *et al.*, Net continuous wave optical gain in a low loss silicon-on-insulator waveguide by stimulated Raman scattering, Opt. Express 13 (2005) 519-525.

[16] H. Rong, *et al.*, Low-threshold continuous-wave Raman silicon laser, Nat Photon 1 (2007) 232-237.

[17] V. Raghunathan, *et al.*, Demonstration of a Mid-infrared silicon Raman amplifier, Opt. Express 15 (2007) 14355-14362.

[18] V. Raghunathan, *et al.*, Nonlinear absorption in silicon and the prospects of mid-infrared silicon Raman lasers, physica status solidi (a) 203 (2006) R38-R40.

[19] H. Rong, *et al.*, A continuous-wave Raman silicon laser, Nature 433 (2005) 725-728.

[20] A. L. H. Rong, R. Jones, 0. Cohen, D. Hak, R. Nicolaescu, A. Fang, and M. Paniccia,, An all-silicon Raman laser, Nature 7023 (2005) 292-294.

[21] O. Boyraz and B. Jalali, Demonstration of a silicon Raman laser, Opt. Express 12 (2004) 5269-5273.

[22] R. Espinola, *et al.*, Raman amplification in ultrasmall silicon-on-insulator wire waveguides, Opt. Express 12 (2004) 3713-3718.

[23] R. Claps, *et al.*, Observation of stimulated Raman amplification in silicon waveguides, Opt. Express 11 (2003) 1731-1739.